\documentclass[12pt]{article}
\usepackage{epsfig}

\def\beq{\begin{equation}}
\def\eeq#1{\label{#1}\end{equation}}
\def\eeqn{\end{equation}}

\def\beqa{\begin{eqnarray}}
\def\eeqa#1{\label{#1}\end{eqnarray}}
\def\eeqan{\end{eqnarray}}

\let\bar=\overbar

\def\Dslash{\not{\hbox{\kern-4pt $D$}}}
\def\dslash{\not{\hbox{\kern-2pt $\del$}}}

\def\msb{{\bar{\ssstyle M \kern -1pt S}}}

\usepackage{fancyhdr,graphicx}
\def\Title#1{\begin{center} {\Large {\bf #1} } \end{center}}

\begin{document}

\Title{Signal of quark deconfinement and thermal evolution of
hybrid stars}

\bigskip\bigskip


\begin{raggedright}

{\it Miao Kang\index{Med, I.}\\
The College of Physics and Electronics\\
Henan University\\
Kaifeng, 475004\\
P. R. China\\
{\tt Email: kangmiao07@gmail.com}}
\bigskip\bigskip
\end{raggedright}

\section{Introduction}

It is well known that neutron stars spin down due to magnetic
dipole radiation. The deconfinement phase transition of hadron
matter to quark matter is expected to occur in the dense cores of
the stars during spin-down~\cite{76bay}\cite{92gle}. The phase
transition continuously takes place inducing not only structural
changes but also energy release in case of a first-order phase
transition. The generation of energy increases the internal energy
of the stars which is called deconfinement
heating~\cite{91hae}\cite{06yu}\cite{07kan}. The temperature of
the stars arise when the deconfinement heating appears in the
cores. We explore the deconfinement signature by studying the
changes of surface temperature in the thermal evolution process of
neutron stars.

\section{Method and Result}
Following Glendenning's hybrid stars
model~\cite{92gle}\cite{97gle}, we use a standard two-phase
description of the equation of state(EOS) through which the hadron
and quark phases are modelled separately. The resulting EOS of the
mixed phase is obtained by imposing Gibbs's
conditions~\cite{00sch}. The Argonne
$V18+\delta\upsilon+UIX^{*}$(APR) model~\cite{98akm} of hadronic
matter and the MIT bag model of quark matter~\cite{97sch} are used
to construct the model of stars. By treating a rotating star as a
perturbation on a non-rotating star and by expanding the metric of
an axially symmetric rotating star in even powers of the angular
velocity $\Omega$, we can obtain the structure of the rotating
stars \cite{67har} as in Kang \& Zheng \cite{09kan}.

The deconfinement heating is coupled with rotating evolution of
the stars. Combining the energy change with the evolutionary
structure of hybrid stars, we get the total heat
luminosity~\cite{07kan}\cite{07mia}
\begin{equation} H_{dec}=\int
\frac{de}{dv}\dot{v}(t)\rho_{B}dV
\end{equation}
 the rotation frequency is
given by
\begin{equation}
\dot{v}=-\frac{16\pi^{2}}{3Ic^{3}}\mu^{2}v^{3}\sin^{2}\theta.
\end{equation}

The traditional standard cooling model often based on the
Tolman-Oppenheimer-Volkoff equation of hydrostatic
equilibrium~\cite{05pag}\cite{04yak} do not consider rotational
evolution of the stars.
 We combine the equation of thermal balance with the rotating structure equations of the
stars and rewrite the energy equation in the approximation of an
isothermal interior(Kang \& Zheng \cite{09kan})
\begin{equation}
C_{V}(T_{i},v)\frac{dT_{i}}{dt}=-L_{\nu}^{\infty}(T_{i},v)-L_{\gamma}^{\infty}(T_{s},v)
\end{equation}
\begin{equation}
C_{V}(T_{i},v)=\int_{0}^{R(v)}c(r,T)(1-\frac{2M(r)}{r})^{-1/2}4\pi
r^{2}dr
\end{equation}
\begin{equation}
 L_{\nu}^{\infty}(T_{i},v)=\int_{0}^{R(v)}\varepsilon(r,T)
(1-\frac{2M(r)}{r})^{-1/2}e^{2\Phi}4\pi r^{2}dr.
\end{equation}
Where $C_{V}(T_{i},v)$ is the total stellar heat capacity,
$L_{\nu}^{\infty}(T_{i},v)$ and $L_{\gamma}^{\infty}=4\pi
R^{2}(v)\sigma T_{s}^{4}(1-R_{g}/R)$ are the total redshifted
neutrino luminosity and the surface photon luminosity,
respectively, which are functions of rotation frequency and
temperature. Neutrino emission is generated in numerous reactions
in the interior of neutron stars, e.g. as reviewed by Page et
al.~\cite{05pag}. The most powerful neutrino emission is provided
by the direct Urca processes(the nucleon direct Urca process and
the quark direct Urca process).  Using to Eqs.(2)-(5), we can
simulate the thermal evolution of neutron stars with deconfinement
heating.

\begin{figure}
   \centering
    \epsfig{file=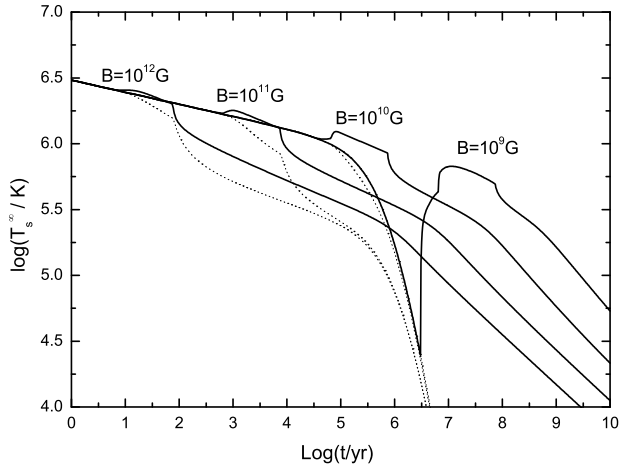,height=3.0in}
   \caption{Thermal evolution curves of a 1.6 $M_{\odot}$ neutron star with
   deconfinement heating
   for various magnetic field strengths(solid curves) and the curves without deconfinement heating (dotted
   curves))
 }
   \label{Fig:f1}
   \end{figure}

In Fig~\ref{Fig:f1}, we present thermal evolution behavior of a
1.6$M_{\odot}$ neutron star
     for different magnetic fields ($10^{9}-10^{12}$G).
Due to the coupling of thermal evolution and spin-down, all
curves(with deconfinement heating and without deconfinement
heating) show clear magnetic field dependence.
 It is evident
     that the temperature of the curves with deconfinement heating(solid curves)
    are higher than for the standard cooling scenario without deconfinement(dotted curves).
    We can observe a competition between cooling
and heating processes from the heating curves, where deconfinement
heating can produce a characteristic
     rise of surface temperature and even dominate the history of thermal evolution. Eventually, they reach a thermal equilibrium,
     where the heat generated is radiated away at the same rate from the star surface. We find the weaker magnetic field have the larger change of
      temperature. The low magnetic field ($10^{9}$G) produces a sharp jump in surface temperature as soon as the deconfinement quark matter
      appearing during spin-down. Intermediate magnetic field ($10^{10},10^{11}$G) lead to slight changes in the temperature, but high magnetic field form
  only the temperature plateau at a time.

\begin{figure}[htb]
  \begin{center}
 \epsfig{file=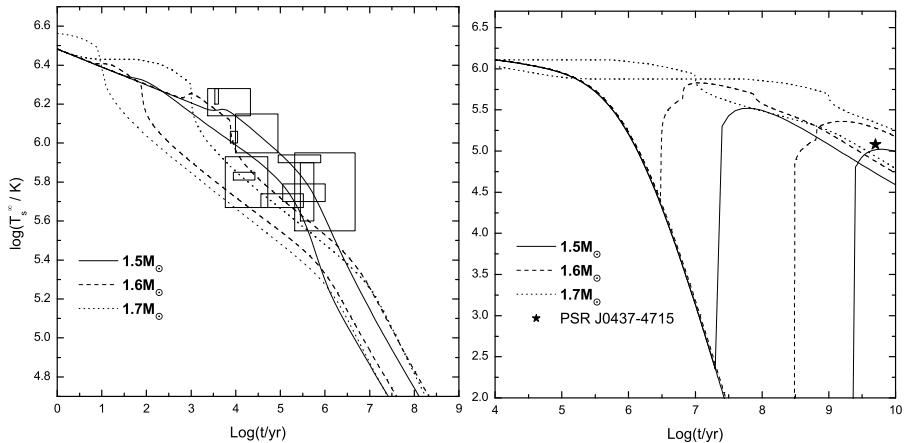,height=3.0in}
   \caption{Thermal evolution curves of neutron stars with deconfinement heating for different stars masses and B=$10^{12}$,$10^{11}$G(left panel) and
   B=$10^{9}$,$10^{8}$G(right panel. Rectangles in the left-hand panel indicate observational data on cooling neutron stars with strong magnetic fields.)
 }
   \label{Fig:f2}
   \end{center}
   \end{figure}

 In Fig~\ref{Fig:f2}, we present the cooling behavior of different
masses stars for magnetic field B=$10^{12},10^{11}$G (left panel)
and magnetic field B=$10^{9},10^{8}$G(right panel) with
deconfinement heating. The observational data, taken from tables 1
and 2 in Page et al.\cite{04pag}, have been shown in left panel.
Comparing with
 previous investigation~\cite{07kan}, we find the
thermal evolution curves of our present work are more compatible
with the observational data(left panel). In our present study,
using the APR EOS, NDU processes can not be triggered easily in
stars which lead to the higher temperatures of the evolution
curves than in the previous cases.
 In the cases of weak magnetic field, stars have
high temperatures($>10^{5}$K) at older ages ($>10^{9}$yrs). We
thus think that
 high temperature of some millisecond pulsars with low magnetic fields~\cite{04kar}, especially for PSR J0437-4715,
  can be explained using the deconfinement heating model of hybrid stars.
   We can observe that 1.5 $M_{\odot}$ stars follow a similar thermal evolution track as 1.6 $M_{\odot}$, but there is not a period increasing in temperature
   for 1.7$M_{\odot}$. The reason for this is that the quark matter to appear at the birth of the stars for 1.7$M_{\odot}$; For
1.5 $M_{\odot}$ and 1.6 $M_{\odot}$ stars, quark deconfinement
occurs when the central density gradually increases during
spin-down, which results in the temperatures of the stars to
increase rapidly. This is a characteristic signal as quark matter
 arises during the rotational spin-down of stars for weak magnetic case.
\section{Conclusions}
Our results show that deconfinement heating can drastically affect
the thermal evolution of neutron stars. The rise of surface
temperature of cooling stars, as a signature of quark
deconfinement, is derived from the deconfinement heating. It is
noteworthy that a significant rise of the temperature accompanies
the appearance of quark matter at older ages for low magnetic
field stars. This may be a evidence for existence of quark matter,
if a period of rapid heating is observed for a very old pulsar.
Deconfinement heating provides a new way to study the signal of
deconfinement.

\section{Acknowledgements}
This work is supported by NFSC under Grant Nos.10747126 and
10773004.







\end{document}